\documentclass[12pt]{article}
\usepackage{subeqn}
\usepackage{epsfig}
\newcommand{\be}{\begin{equation}}
\newcommand{\ee}{\end{equation}}

\newcommand{\no}{\noindent}
\newcommand{\ce}{\begin{center}}
\newcommand{\nc}{\end{center}}

\makeatletter
\@addtoreset{equation}{section}
\makeatother

\def\sqr#1#2{{\vcenter{\vbox{\hrule height.#2pt
\hbox{\vrule width.#2pt height#1pt \kern#1pt
\vrule width.#2pt} \hrule height.#2pt}}}}

\def\operp{\hbox{${\kern+.25em{\bigcirc}
\kern-.85em\bot\kern+.85em\kern-.25em}$}}

\def\lsim{\;\raise0.3ex\hbox{$<$\kern-0.75em\raise-1.1ex\hbox{$\sim$}}\;}
\def\gsim{\;\raise0.3ex\hbox{$>$\kern-0.75em\raise-1.1ex\hbox{$\sim$}}\;}
\def\no{\noindent}

\def\ce{\centerline}
\def\ve{\vfill\eject}
\def\rdots{\mathinner{\mkern1mu\raise1pt\vbox{\kern7pt\hbox{.}}\mkern2mu
\raise4pt\hbox{.}\mkern2mu\raise7pt\hbox{.}\mkern1mu}}

\def\e e{$e^+ e^-$ }

\begin{document}

\ce{\bf SCALAR TREFOILS}
\vskip.5cm

\ce{\it Robert J. Finkelstein}
\vskip.3cm

\ce{Department of Physics and Astronomy}
\ce{University of California, Los Angeles, CA 90095-1547}
\vskip1.0cm

\no {\bf Abstract.}  The knot model is extended by assuming that the trefoils
are realized as either chiral fermions or as scalar bosons.  There are then
four scalar trefoils with electric charges $(0,-1,2/3,-1/3)$ that may be
classified in the same way as the chiral fermions: as two isotopic 
doublets where the two doublets have different hypercharge and the two
members of the doublets have different $t_3$.  Only the neutral scalar
plays the role of the standard Higgs in fixing the mass ratios of the
vector bosons, while the charged scalars, in addition to having the usual
electromagnetic interactions of scalar particles, fix the mass spectrum
of the fermions.  The extended model would suggest a search for the
charged scalars.

\vskip4.5cm

\no UCLA/07/TEP/18

\ve

\section{Introduction.}

In view of the existence and topological stability of knots in classical
field theory, it is natural to explore the role of the same structures in
quantum field theory and in particular to examine the possibility of 
interpreting the field quanta of the standard theory as knotted solitons
instead of point particles.  
When the standard point particles are replaced by the knots, one finds that the topological structure
of the fermionic knot is determined by its electric charge, hypercharge,
and isotopic spin.${}^{1,2,3}$  This result permits a geometric interpretation of 
charge, i-spin, and hypercharge and thereby also permits one to
assign a knot structure to the vector bosons.  Among other consequences it then also becomes natural to expand the Higgs sector so that there is a multiplet of 4 scalars, instead of a single Higgs, with the same charges
as the fermions.

\vskip.5cm

\section{The Knot Conjecture.}

If the elementary particles are quantum knots, one expects that the simplest
knots (trefoils) are realized as the most
elementary particles (chiral fermions and scalar bosons). 
To test this idea we may start from the observation that there
are 4 trefoils and 4 families of elementary fermions (neutrinos, leptons,
up quarks, down quarks).  We shall assume that the 3 members of each
family $(e,\mu,\tau),~(\nu_e,\nu_\mu,\nu_\tau),~(u,c,t),~(d,s,b)$ are the
three lowest states of the 4 fermionic solitons.  

We shall now also assume
that there are 4 scalar knots as well as 4 spinor knots.  Then the complete
wave function of the spinor or the scalar soliton is the product of the 
standard chiral spinor or scalar wave function multiplied by a knot factor.  The
knot factor depends on the symmetry algebra of the knot, which is
$SU_q(2)$, and we take the quantum state of the knot to be determined by an
irreducible representation of $SU_q(2)$, denoted by $D^j_{mm^\prime}$.  We
shall now assume that the knot factor is the quantum state of the knot,${}^3$ namely
\be
D^j_{mm^\prime} = D^{N/2}_{\frac{w}{2}\frac{r+1}{2}} \qquad
\mbox{or equivalently} \qquad D^{3t}_{-3t_3-3t_0}
\ee
with

\ve

\begin{eqnarray}
N &=& 6t \nonumber \\
w &=& -6t_3 \\
r+1 &=& -6t_0 \nonumber
\end{eqnarray}
where $D^j_{mm^\prime}$ is a matrix element of the $2j+1$ irreducible
representation of the quantum group $SU_q(2)$.  Here $N,w$, and $r$ are
the number of crossings, the writhe, and the rotation of the knot 
while $t,t_3$, and
$t_0$ are i-spin, 3-component of the i-spin, and hypercharge.  For the
spinor and scalar realization of the trefoil one has $N=3$ while the 4
separate realizations of the trefoil are described in Table 1.
\[
\begin{array}{c|crrrrrc}
& t & t_3 & t_0 & w & r & Q & D^{3/2}_{mm^\prime} \\
\hline
\nu & \frac{1}{2} & \frac{1}{2} & -\frac{1}{2} & -3 & 2 & 0 & 
D^{3/2}_{-\frac{3}{2}+\frac{3}{2}} \\
e^- & \frac{1}{2} & -\frac{1}{2} & -\frac{1}{2} & 3 & 2 & -e & 
D^{3/2}_{\frac{3}{2}\frac{3}{2}} \\
u & \frac{1}{2} & \frac{1}{2} & \frac{1}{6} & -3 & -2 & \frac{2}{3}e & 
D^{3/2}_{-\frac{3}{2}-\frac{1}{2}} \\
d & \frac{1}{2} & -\frac{1}{2} & \frac{1}{6} & 3 & -2 & -\frac{1}{3}e & 
D^{3/2}_{\frac{3}{2}-\frac{1}{2}}
\end{array}
\]
\ce{\bf Table 1.}
\vskip.3cm

In the table the first three columns describe the standard description of
the four classes of fermions $(\nu,e,u,d)$ and the next three columns
describe the writhe, rotation and electric charge of the same classes
regarded as knots.  In the last column is the corresponding irreducible
representation where $m = \frac{w}{2}$ and $m^\prime = \frac{r+1}{2}$.  
For a detailed justification of this table and Eq. 
(2.2), see Ref. 3.  We are now
assuming that there is a single table for scalars and chiral fermions.

By (2.1) and (2.2) one has Table 2 for the 4 vector bosons.

\ve

\[
\begin{array}{c|rrcc} 
& t & t_3 & t_0 & D^j_{mm^\prime} \\
\hline
W^+ & 1 & 1 & 0 & D^3_{-30} \\
W^- & 1 & -1 & 0 & D^3_{30} \\
W^3 & 1 & 0 & 0 & D^3_{00} \\
W^0 & 0 & 0 & 0 & D^0_{00}
\end{array}
\]
\ce{\bf Table 2.}
\vskip.3cm

\no We shall abbreviate the knot factor (the quantum state of the knot) as
follows:
\begin{eqnarray}
\mbox{For spinor and scalar fields write}~ D_k \quad & &k = (\nu,e,u,d) \\
\mbox{For vector bosons write}~ {\cal{D}}_\alpha 
\hskip2.1cm & &\alpha = (+,-,3,0)
\end{eqnarray}

Since the $D^j_{mm^\prime}$ are irreducible representations of $SU_q(2)$
we shall next collect the relevant information about $SU_q(2)$.

\vskip.5cm

\section{The Knot Algebra.}

One way of seeing that $SL_q(2)$ is the appropriate algebra of the
knot is to observe on the one hand 
that the Kauffman algorithm (for generating the Kauffman or the
Jones polynomial that characterizes a knot) may be expressed
in terms of the matrix
\be
\epsilon_q = \left(
\begin{array}{cc}
0 & q^{-1/2} \\
-q^{1/2} & 0 
\end{array} \right) \qquad
\epsilon_q^2 = -1
\ee
and on the other hand that $\epsilon_q$ is also the invariant matrix of $SL_q(2)$ since
\be
T^t\epsilon_q T = T\epsilon_q T^t = \epsilon_q
\ee
where $T$ belongs to a two-dimensional representation of
$SL_q(2)$. 

We shall now describe this algebra.  Let 
\be
T = \left(
\begin{array}{cc}
a & b \\ c & d
\end{array} \right)
\ee

Then by (3.2) the matrix elements of $T$ satisfy the following algebra
\begin{center}
\begin{tabular}{llll}
$ab = qba$ \quad & $bd = qdb$ \quad & $ad-qbc = 1$ \quad & $bc = cb$ \\
$ac = qca$ \quad & $cd = qdc$ \quad & $da-q_1cb = 1$ \quad & $q_1 = q^{-1}$   \hskip2.5cm (A)
\end{tabular}
\end{center}

In the discussion of electroweak we need only the unitary
subalgebra obtained by setting
\begin{eqnarray*}
d &=& \bar a \\
c &=& -q_1\bar b
\end{eqnarray*}
Then $(A)$ reduces to the following
\begin{center}
\begin{tabular}{lll}
$ab = qba$ \qquad & $a\bar a + b\bar b = 1$ \qquad & $b\bar b = 
\bar bb$ \\
$a\bar b = q\bar ba$ \qquad & $\bar aa + q_1^2\bar bb = 1$ \qquad &
\hskip4.5cm $(A)^\prime$
\end{tabular}
\end{center}
For the physical applications we need the higher representations of
$SU_q(2)$.  The $2j+1$-dimensional unitary irreducible
representations of the $SU_q(2)$ algebra $(A)^\prime$ are
\be
D^j_{mm^\prime} = \sum_{s,t} A^j_{mm^\prime}(s,t)
\delta(s+t,n_+^\prime)a^sb^{n_+-s}\bar b^t\bar a^{n_--t}
\ee
where
\[
A^j_{mm^\prime}(s,t) = \left[{\langle n^\prime_+\rangle_1!~
\langle n^\prime_-\rangle_1!\over
\langle n_+\rangle_1!~\langle n_-\rangle_1!}\right]^{1/2} 
\left\langle\matrix{n_+\cr s}\right\rangle_1~
\left\langle\matrix{n_-\cr t}\right\rangle_1~q^{t(n_+-s+1)}
(-)^t 
\]
and
\[
\begin{array}{rcl}
n_\pm &=& j\pm m \\ n^\prime_\pm &=& j\pm m^\prime \\
\end{array} \quad
\left\langle\matrix{n \cr s}\right\rangle_1 =
{\langle n\rangle_1!\over \langle s\rangle_1!\langle n-s\rangle_1!}
\quad \langle n\rangle_1 = {q_1^{2n}-1\over q_1^2-1} 
\]

The algebra $(A)^\prime$ is invariant under the gauge transformations
\be
\begin{array}{rcl}
a^\prime  = e^{i\varphi_a}a \hskip1.0cm  & & 
b^\prime = e^{i\varphi_b}b \\
\bar a^\prime = e^{-i\varphi_a}\bar a \hskip0.90cm & &
\bar b^\prime = e^{-i\varphi_b}\bar b \\
\end{array}
\ee
These transformations induce the following gauge transformations on
$D^j_{mm^\prime}$.${}^3$
\be
D^{j~~\prime}_{mm^\prime} = U_aU_bD^j_{mm^\prime}
\ee
where
\be
\begin{array}{rcl}
U_a &=& e^{i\varphi_a(m+m^\prime)} \\
U_b &=& e^{i\varphi_b(m-m^\prime)} \\
\end{array}
\ee
and by (2.1) and (2.2)
\be
\begin{array}{rcl}
m+m^\prime &=& -\frac{1}{2}~(w+r+1) = -3(t_3+t_0) \\
m-m^\prime &=& -\frac{1}{2}~(w-r-1) = -3(t_3-t_0) \\
\end{array}
\ee

Since $b$ and $\bar b$ commute, they have common eigenstates.  Let 
$|0\rangle$ be designated as a ground state and let
\begin{eqnarray}
b|0\rangle &=& \beta|0\rangle \\
\bar b|0\rangle &=& \beta^\star|0\rangle
\end{eqnarray}
and
\be
\bar bb|0\rangle = |\beta|^2|0\rangle
\ee
where $\bar bb$ is Hermitian with real eigenvalues and orthogonal
eigenstates.

One finds by $(A)^\prime$ that
\be
\bar bb|n\rangle = E_n|n\rangle
\ee
where
\be
|n\rangle \sim\bar a^n|0\rangle
\ee
and
\be
E_n = q^{2n}|\beta|^2
\ee
$\bar bb$ resembles the Hamiltonian of an oscillator but with eigenvalues
arranged in geometrical progression and with $|\beta|^2$ corresponding to
$\frac{1}{2}~\hbar\omega$.  If we take $H(\bar bb)$ to be the Hamiltonian of
the knot, where the functional form of $H$ is left unspecified, it will have
the same eigenstates.

\vskip.5cm

\section{The Normal Modes of the Quantum Fields.}

We assume that the separate fermion and scalar states are the low lying
states of the fermion and scalar solitons as follows:
\be
D_k|n\rangle \qquad k = (\nu,\ell,u,d),~~n = (0,1,2)
\ee
by (2.3), and
where $|n\rangle$ is the $n^{\rm th}$ level of the ``$q$-oscillator"
given by (3.13).  
Then
the complete $L$-chiral normal modes of the Dirac fields are
\be
L_{kn} = \chi_{kn}(\vec p,\vec t) D_k|n\rangle
\ee

The corresponding normal modes of the scalar fields are
\be
\Phi_{kn} = \varphi_{kn}(\vec p,\vec t) D_k|n\rangle
\ee
Here $\chi_{kn}$ and $\varphi_{kn}$ are normal modes of the standard
Dirac and of the hypothetical scalar fields, respectively.

The standard theory is invariant under local $SU(2)$ and the fermions and
scalars are doublets under $SU(2)$.  The two members of the doublet differ
in $t_3$ (or in the writhe of the knot), while the $(\nu,\ell)$ doublets
differ from the $(u,d)$ doublets in the hypercharge (or in the rotation of
the knot).  Therefore we have the two doublet solitons 
differing in hypercharge (or rotation).
\be
\left(
\begin{array}{c}
\chi_\nu D_\nu \\ \chi_\ell D_\ell
\end{array} \right) \quad \mbox{and} \quad \left(
\begin{array}{c}
\chi_u D_u \\ \chi_d D_d
\end{array} \right)
\ee

The vector connection in the standard theory is
\be
W_+t_+ + W_-t_- + W_3t_3 + W_0t_0
\ee
where $(t_+,t_-,t_3,t_0)$ are the generators of the standard electroweak
theory in the charge representation and
\be
t_+ = \left(
\begin{array}{cc}
0 & 1 \\ 0 & 0
\end{array} \right)~,~~t_- = \left(
\begin{array}{cc}
0 & 0 \\ 1 & 0 
\end{array} \right)~,~~t_3 = \left(
\begin{array}{cc}
1 & 0 \\ 0 & -1
\end{array} \right)~,~~t_0 = \left(
\begin{array}{cc}
1 & 0 \\ 0 & 1
\end{array} \right)
\ee

We now replace (4.5) by
\be
W_+\tau_+ + W_-\tau_- + W_3\tau_3 + W_0\tau_0
\ee
where
\be
\tau_k = c_k(q,\beta)t_k{\cal{D}}_k \qquad k = (+,-,3,0)
\ee
and the ${\cal{D}}_k$ are the charge states of the four vector mesons
\begin{eqnarray}
{\cal{D}}_+ &\equiv& D^3_{-30}/N_+ = \bar b^3\bar a^3 \\
{\cal{D}}_- &\equiv& D^3_{03}/N_- = a^3b^3 \\
{\cal{D}}_3 &\equiv& D^3_{00} = f_3(\bar bb) \\
{\cal{D}}_0 &\equiv& D^0_{00} = 1
\end{eqnarray}
where $f_3(\bar bb)$ is computed from (3.4) and
where $N_+$ and $N_-$ are numerical factors following from (3.4).

The $c_k$ are numerical functions of the parameters $(q,\beta)$ and are
partially fixed by relations between the masses of the vector bosons.  The
$c_k$ are also proportional to the coupling constants and these are
different for $\vec W$ and $W_0$.

For the vector connection we shall set
\be
{\cal{W}}_\mu = W_\mu^\alpha\tau_\alpha
\ee
Then the covariant derivative is
\be
\nabla = 1~\partial + {\cal{W}}
\ee

\vskip.5cm

\section{Quantum Field Theory.}

The quantum field theory is defined by a Lagrangian invariant under a
group of local gauge transformations and by boundary conditions defined by
the ground states of the fields.  The fields and states are defined only up
to a gauge transformation.  To select the ground state from the ensemble
of gauge equivalent ground states we select a special gauge.  This
privileged gauge, along with both the Fock vacuum and the lowest state
of the $q$-algebra, limits the ground state of the model which will now be
denoted by $|0\rangle\rangle$,

We shall further define the ground state by the requirement that all the
interacting fields be independent of position in this state in this
state.  Then all kinetic energy terms in this state will vanish.  The constant fields shall also be
chosen to minimize the potential so that the total energy will be
minimized for this vacuum state.

Now define the Lorentz invariant scalar products 
$\stackrel{\bar\circ}{\varphi_k}\stackrel{\circ}{\varphi_k}$ and
$\stackrel{\bar\circ}{W_\alpha^\mu} \stackrel{\circ}{W_{\alpha\mu}}$ by
\begin{eqnarray}
& &\stackrel{\bar\circ}{\varphi_k}\stackrel{\circ}{\varphi_k} = \langle
\!\langle 0|\bar\Phi_k
\Phi_k|0\rangle\!\rangle \hskip 2.5cm  k = \nu,\ell,u,d \\
& &\stackrel{\bar\circ}{W_\alpha^\mu}\stackrel{\circ}{W_{\alpha\mu}} =
\langle\!\langle 0|\bar W_\alpha^\mu W_{\alpha\mu}
|0\rangle\!\rangle \hskip1.5cm
\alpha = (+,-,3,0)
\end{eqnarray}
where $\Phi_k$ and $W_\alpha^\mu$ are the position independent fields
defined over the ground state and both sides of these equations are to be
taken in the privileged gauge.  In the Goldstone-Higgs language the
Goldstone fields, which are parameters of the gauge transformation, are
gauged to vanish in the privileged gauge.  In the same gauge the 
ground states of four scalar solitons are arranged in two doublets that take the following
form:
\be
\left(
\begin{array}{c}
\phi_\nu D_\nu|0\rangle \\
\phi_\ell D_\ell|0\rangle
\end{array} \right) \quad \mbox{and} \quad \left(
\begin{array}{c}
\phi_u D_u|0\rangle \\
\phi_d D_d|0\rangle
\end{array} \right)
\ee
labelled by hypercharge $t_0 = -1/2$ and $t_0 = 1/6$ (or by rotation
$r=2$ and $r=-2$) respectively.  The three Goldstone degrees of freedom
associated with $SU(2)$ gauge transformations are expressed not as Goldstone bosons, but as the three degrees
of freedom associated with the longitudinal modes of the three massive
vectors in the standard way.

\vskip.5cm

\section{Masses of the Vectors.}

The invariant form of the kinetic energy of each scalar doublet is
\be
\frac{1}{2}~ \mbox{Tr}~ {\overline{\nabla_\mu}}\Phi_k \nabla^\mu\Phi_k
\ee
where $k$ signifies the hypercharge (or the rotation of the knot).  Here
\be
\nabla_\mu = \partial_\mu + W_\mu^\alpha\tau_\alpha
\ee
and $\Phi_k$ is one of the doublets described by (5.3).

Let us write
\be
\Phi_k = \left(
\begin{array}{c}
u_k \\ v_k
\end{array} \right)
\ee
and let us consider $k = t_0 = -1/2$.  Then
\be
\Phi(t_0 = -1/2) = \left(
\begin{array}{c}
\varphi_\nu D_\nu|0\rangle \\ \varphi_\ell D_\ell|0\rangle
\end{array} \right)
\ee
and
\begin{eqnarray}
\nabla_\mu\Phi\left(-\frac{1}{2}\right) = \partial_\mu\left(
\begin{array}{c}
\varphi_\nu D_\nu|0\rangle \\ \varphi_\ell D_\ell|0\rangle
\end{array} \right) &+& \left(
\begin{array}{c}
W_\mu^+{\cal{D}}_+\varphi_\ell D_\ell|0\rangle \\ 0
\end{array} \right) + \left(
\begin{array}{c}
0 \\ W_\mu^-{\cal{D}}_-\varphi_\nu D_\nu|0\rangle
\end{array} \right) \nonumber \\
&+&{\rm neutral~couplings}\left(
\begin{array}{c}
\varphi_\nu D_\nu|0\rangle \\ \varphi_\ell D_\ell|0\rangle
\end{array} \right)
\end{eqnarray}
The neutral couplings, written in terms of the physical fields, $A$ and
$Z$, are
\be
i~gW_\mu^3\tau_3+ig_0W_\mu^0\tau_0 = {\cal{A}}A_\mu +
{\cal{Z}}Z_\mu
\ee
where
\begin{eqnarray}
{\cal{A}} &=& i(g\tau_3\sin\theta + g_0\tau_0\cos\theta) = 
ie(\tau_3+\tau_0) \\
{\cal{Z}} &=& i(g\tau_3\cos\theta-g_0\tau_0\sin\theta) = ie
(\cot\theta~\tau_3-\tan\theta~\tau_0)
\end{eqnarray}
and $\theta$ is the Weinberg angle defined by
\be
\sin\theta = \frac{e}{g} \quad \mbox{or} \quad \cos\theta = \frac{e}{g_0}
\ee
Note that
\be
{\cal{A}}|\nu\rangle = 0
\ee
if $|\nu\rangle$ is any neutral state and therefore by (6.7)
\be
(\tau_0+\tau_3)|\nu\rangle = 0
\ee
Then by (6.8) and the preceding equation
\[
{\cal{Z}}|\nu\rangle = ie(\cot\theta + \tan\theta)\tau_3|\nu\rangle
\]
or
\be
{\cal{Z}}|\nu\rangle = \frac{ig}{\cos\theta}~\tau_3|\nu\rangle
\ee
Then by (6.6)-(6.8)
\be
\mbox{[neutral~couplings]}\left(
\begin{array}{c}
\varphi_\nu D_\nu|0\rangle \\
\varphi_\ell D_\ell|0\rangle
\end{array} \right) = \left(
\begin{array}{c}
0 \\ {\cal{A}}_\ell\varphi_\ell D_\ell|0\rangle
\end{array} \right)~A_\mu + \left(
\begin{array}{c}
{\cal{Z}}_\nu\varphi_\nu D_\nu|0\rangle \\
{\cal{Z}}_\ell\varphi_\ell D_\ell|0\rangle
\end{array} \right) Z_\mu
\ee
where by (4.8)
\[
{\cal{A}}_\ell|\ell\rangle = ie(c_3t_3{\cal{D}}_3+c_0t_0{\cal{D}}_0)|\ell\rangle
\]
The kinetic energy (6.1) may now be written as
\be
\begin{array}{rcl}
& &\mbox{}\langle 0|\bar D_\nu\partial_\mu\bar\varphi_\nu\cdot\partial^\mu
\varphi_\nu D_\nu|0\rangle + \langle 0|\bar D_\ell\partial_\mu\bar\varphi_\ell
\cdot \partial^\mu\varphi_\ell D_\ell|0\rangle \\
&+&\langle 0|\bar D_\ell\bar{\cal{D}}_+{\cal{D}}_+ D_\ell|0\rangle
\bar\varphi_\ell\varphi_\ell \bar W^{+\mu}W^+_\mu +
\langle 0|\bar D_\nu\bar{\cal{D}}_-{\cal{D}}_-D_\nu|0\rangle
\bar\varphi_\nu\varphi_\nu \bar W^{-\mu}W_\mu^- \\
&+&\langle 0|\bar D_\ell\bar {\cal{A}}_\ell {\cal{A}}_\ell D_\ell|0\rangle\bar\varphi_\ell\varphi_\ell
A^\mu A_\mu + \langle 0|\bar D_\nu\bar{\cal{Z}}_\nu{\cal{Z}}_\nu
D_\nu|0\rangle\bar\varphi_\nu\varphi_\nu Z^\mu Z_\mu \\
&+& \hskip2.0cm 0 \hskip2.5cm + 
\langle 0|\bar D_\ell\bar{\cal{Z}}_\ell{\cal{Z}}_\ell D_\ell|0\rangle
\bar\varphi_\ell\varphi_\ell Z^\mu Z_\mu
\end{array}
\ee
The part of the kinetic energy contributing to the total Lagrangian and
coming from the $\nu$-scalar is
\be
I_\nu\partial_\mu\bar\varphi_\nu\partial^\mu\varphi_\nu +
I_{\nu--}\bar\varphi_\nu\varphi_\nu(W_1^\mu W_{1\mu} + W_2^\mu W_{2\mu})
+ I_{\nu{\cal{Z}}{\cal{Z}}}\bar\varphi_\nu\varphi_\nu
Z^\mu Z_\mu
\ee
where
\begin{eqnarray}
I_\nu~~~~&=& \frac{1}{2}~{\rm Tr}~\langle 0|\bar D_\nu D_\nu|0\rangle \\
I_{\nu--}~&=& \frac{1}{2}~{\rm Tr}~\langle 0|\bar D_\nu\bar{\cal{D}}_-
{\cal{D}}_- D_\nu|0\rangle \\
I_{\nu{\cal{Z}}{\cal{Z}}}~ &=& \frac{1}{2}~{\rm Tr}~
\langle 0|\bar D_\nu\bar{\cal{Z}}_\nu{\cal{Z}}_\nu D_\nu|0\rangle
\end{eqnarray}
The part of the kinetic energy contributing to the total Lagrangian and
coming from the $\ell$-scalar is
\be
\begin{array}{rcl}
& &\mbox{} I_\ell\partial_\mu\bar\varphi_\ell\partial^\mu\varphi_\ell +
I_{\ell ++}(W_1^\mu W_{1\mu}+W_2^\mu W_{2\mu})\bar\varphi_\ell
\varphi_\ell \\
& &\mbox{}+ I_{\ell aa}\bar\varphi_\ell\varphi_\ell A^\mu A_\mu +
I_{\ell{\cal{Z}}{\cal{Z}}}\bar\varphi_\ell\varphi_\ell Z^\mu Z_\mu
\end{array}
\ee
where
\begin{eqnarray}
I_\ell~~ &=& \frac{1}{2}~{\rm Tr}~\langle 0|\bar D_\ell D_\ell|0\rangle \\
I_{\ell ++} &=& \frac{1}{2}~{\rm Tr}~
\langle 0|\bar D_\ell\bar{\cal{D}}_+{\cal{D}}_+D_\ell|0\rangle \\
I_{\ell aa}~ &=& \frac{1}{2}~{\rm Tr}~\langle 0|\bar D_\ell\bar{\cal{A}}_\ell
{\cal{A}}_\ell D_\ell|0\rangle \\
I_{\ell{\cal{Z}}{\cal{Z}}}~&=& \frac{1}{2}~{\rm Tr}~
\langle 0|\bar D_\ell\bar{\cal{Z}}_\ell {\cal{Z}}_\ell D_\ell|0\rangle
\end{eqnarray}
In interpreting these terms set
\begin{eqnarray}
\varphi_\ell &=& \stackrel{\circ}{\varphi_\ell} + u_\ell \\
\varphi_\nu &=& \stackrel{\circ}{\varphi_\nu} + u_\nu
\end{eqnarray}
where $\stackrel{\circ}{\varphi_\ell}$ and $\stackrel{\circ}{\varphi_\nu}$
are the vacuum expectation values of $\varphi_\ell$ and $\varphi_\nu$
defined by (5.1).

Now we make the basic assumptions
\begin{eqnarray}
\stackrel{\circ}{\varphi_\ell} &=& 0 \\
\stackrel{\circ}{\varphi_\nu} &=& \rho_\nu \not= 0 \qquad
\bar\rho_\nu = \rho_\nu
\end{eqnarray}
The effective Lagangian of the $\nu$-scalar coming from (6.15) is
\be
I_\nu\partial_\mu\bar u_\nu\partial^\mu u_\nu
\ee
while $W^+$, $W^-$, and $Z$ obtain masses from the following terms of
(6.15).
\be
I_{\nu--}\rho_\nu^2(W_1^\mu W_{1\mu} + W_2^\mu W_{2\mu}) +
I_{\nu{\cal{Z}}{\cal{Z}}}\rho_\nu^2 Z_\mu Z^\mu
\ee

The effective Lagrangian of the $\ell$-scalar is on the other hand by (6.19)
\be
\begin{array}{rcl}
I_\ell\partial_\mu\bar u_\ell\partial^\mu u_\ell &+& 
I_{\ell++}(W_1^\mu W_{1\mu} + W_2^\mu W_{2\mu})\bar u_\ell u_\ell \\
&+& I_{\ell aa}\bar u_\ell u_\ell A_\mu A^\mu + I_{\ell {\cal{Z}}{\cal{Z}}}
\bar u_\ell u_\ell Z_\mu Z^\mu
\end{array}
\ee

The terms in (6.29) that contribute to the $W^+,W^-$ and $Z$ mass now
appear in (6.30) as interaction terms between the same vector fields and
the $\ell$-scalar.  

In discussing the $t_0 = \frac{1}{6}$ doublet we
assume as in (6.26)
\be
\stackrel{\circ}{\varphi_u} = \stackrel{\circ}{\varphi_d} = 0
\ee
and therefore obtain a result analogous to (6.30) rather than (6.28) and
(6.29).

If we had assumed $\stackrel{\circ}{\varphi_\ell} \not= 0$, then 
the vacuum
state would have been charged and the photon field would have had a mass
by (6.30) with $u_\ell$ replaced by $\stackrel{\circ}{\varphi_\ell}$ .  The same remark requires Eq. (6.31).

According to the preceding assumptions only the neutral member of the
trefoil scalar quadruplet supplies mass to the vector bosons.  These
masses depend on the coefficients $I_\nu, I_{\nu --}$ and $I_{\nu{\cal{Z}}
{\cal{Z}}}$ that are given by (6.16)-(6.18), and are functions of $q$ and
$\beta$.  The numerical values of $q$ and $\beta$, and therefore 
the $c_k$, defined in (4.8), are limited by 
the empirical ratios of the masses of the vector
bosons (as discussed in Ref. 2).

\vskip.5cm

\section{Masses of the Fermions.}

According to the standard theory the fermion mass term is
\be
{\cal{M}} = \frac{m}{\rho} (\bar\psi^{\rm L}\varphi\psi^{\rm R} +
\bar\psi^{\rm R}\bar\varphi\psi^{\rm L})
\ee
where $\rho$ is the vacuum expectation value of the Higgs scalar and both
$\psi^{\rm R}$ and the product $\bar\psi^{\rm L}\varphi$ are SU(2)
singlets.  The $\varphi$ doublet is composed of a neutral and a charged
component.  To examine the mass in the standard theory one chooses the
unitary gauge in which the charged component is rotated away.  In order to
represent the masses of the individual fermions a numerical mass matrix
is then introduced.

In the present model we consider the following modification of (7.1).
\be
{\cal{M}}_n = \sum_{t_0}(\bar\psi_n^{\rm L}(t_0)
\Phi_n(t_0)\psi^{\rm R}_n(t_0) +
\bar\psi^{\rm R}_n(t_0)\bar\Phi_n(t_0)\psi_n^{\rm L}(t_0))
\ee
where $\Phi_n(t_0)$ is an SU(2) doublet and $\bar\psi^{\rm L}\Phi(t_0)$
is again a SU(2) singlet to maintain SU(2) gauge invariance.  The sum is
over the two values, (-1/2) and (1/6), of $t_0$.

Here $t_0$ is the hypercharge and we define
\begin{eqnarray}
\Phi_n\left(-\frac{1}{2}\right) &=& \left(
\begin{array}{c}
\rho_\nu D_\nu|n\rangle \\ \rho_\ell D_\ell|n\rangle
\end{array} \right) \\
\Phi_n\left(\frac{1}{6}\right) &=& \left(
\begin{array}{c}
\rho_uD_u|n\rangle \\ \rho_dD_d|n\rangle
\end{array} \right)
\end{eqnarray}
and
\begin{eqnarray}
\psi_n^{\rm L}\left(-\frac{1}{2}\right) &=& \left(
\begin{array}{c}
\chi_\nu D_\nu|n\rangle \\ \chi_\ell D_\ell|n\rangle
\end{array} \right) \\
\psi_n^{\rm L}\left(\frac{1}{6}\right) &=& \left(
\begin{array}{c}
\chi_uD_u|n\rangle \\ \chi_dD_d|n\rangle
\end{array} \right)
\end{eqnarray}
and
\be
\bar\rho_k = \rho_k  \qquad k = \nu,\ell,u,d
\ee
Here $\rho_k$ is the vacuum expectation value of $\varphi_k$ if $k=\nu$ 
but not if $k = (\ell,u,d)$.  In the latter cases the $\rho_k$ 
like the $c_k$ in (4.8) are to be
understood as numerical 
scaling parameters that may be required
to fix the relative masses of the four
trefoils.  

Note that the neutral scalar now plays a role in fixing the masses of the
neutral fermions as well as the masses of the neutral bosons.  On the
other hand the charged scalars play quite different roles in the two
cases.  In the bosonic case the charged scalars are spacetime fields
that play no role in determining the bosonic masses.  In the fermion
mass formula (7.2), however, they do not enter as spacetime fields but
as their ``internal wave functions", the trefoil factors,
$D^{3/2}_{\frac{m}{2}\frac{r+1}{2}}$, lying in the knot algebra.

With these assumptions let us expand the $t_0=-\frac{1}{2}$
part of (7.2) as follows:
\begin{eqnarray}
{\cal{M}}_n\left(-\frac{1}{2}\right) &=&
\langle n|\bar D_\nu\chi_\nu^{\rm L}(n)\rho_\nu D_\nu +
\bar D_\ell\bar\chi_\ell^{\rm L}(n)\rho_\ell D_\ell|n\rangle
\chi^{\rm R} + \mbox{adjoint} \\
&=& \rho_\nu\langle n|\bar D_\nu D_\nu|n\rangle
(\bar\chi_\nu^{\rm L}(n)\chi^{\rm R} + \bar\chi^{\rm R}
\chi_\nu^{\rm L}(n)) \nonumber \\
& &\mbox{}+\rho_\ell\langle n|\bar D_\ell D_\ell|n\rangle
(\bar\chi_\ell^{\rm L}(n)\chi^{\rm R} + \bar\chi^{\rm R}
\chi_\ell^{\rm L}(n))
\end{eqnarray}
$\chi_\ell(n)$ is the left chiral component of $\psi_\ell(n)$
\be
\chi^{\rm L}(n) = \frac{1}{2} (1-\gamma_5)\psi_\ell(n)
\ee
$\chi^{\rm R}$ transforms like the right chiral component of a Dirac
spinor, which we may now take to be $\psi_\ell(n)$:
\be
\chi^{\rm R}(n) = \frac{1}{2}~(1+\gamma_5)\psi_\ell(n)
\ee

Then by (7.9)
\be
{\cal{M}}_n\left(-\frac{1}{2}\right) =
m_\nu(n)\bar\chi_\nu(n)\chi_\nu(n) + m_\ell(n)\bar\chi_\ell(n)\chi_\ell(n)
\ee
where
\begin{eqnarray}
m_\nu(n) &=& \rho_\nu\langle n|\bar D_\nu D_\nu|n\rangle \\
m_\ell(n) &=& \rho_\ell\langle n|\bar D_\ell D_\ell|n\rangle
\end{eqnarray}
One obtains similar results for the quark doublet:
\begin{eqnarray}
m_u(n) &=& \rho_u\langle n|\bar D_uD_u|n\rangle \\
m_d(n) &=& \rho_d\langle n|\bar D_dD_d|n\rangle
\end{eqnarray}
These results may be summarized in the following expressions for the
masses of the elementary fermions
\be
M_n(w,r) = \rho(w,r)\langle n|\bar D^{3/2}_{\frac{w}{2}\frac{r+1}{2}}
D^{3/2}_{\frac{w}{2}\frac{r+1}{2}}|n\rangle
\ee
or
\be
M_n(t_3,t_0) = \rho(t_3,t_0)\langle n|\bar D^{3/2}_{-3t_3~-3t_0}
D^{3/2}_{-3t_3~-3t_0}|n\rangle
\ee
These relations have been discussed in Refs. 1 and 2.  Again the values of
$q$ and $\beta$ need to be adjusted to agree with the empirical mass ratios,
as illustrated in Refs. 1 and 2.

Since there are two parameters, $q$ and $\beta$, and since there appear
to be three masses in each family, ratios of the masses may be computed
without ambiguity.  To determine absolute masses one needs the prefactor
$\rho$ $(t_3,t_0)$.

If there are in fact only three masses in each family, the present model
needs to be supplemented by an exclusion principle allowing only three
masses.

\vskip.5cm

\section{Remarks.}

The starting point of the considerations leading to a knot model of the
elementary particles is the fact that knotted field configurations
appear in certain classical field theories and that they are also topologically stable.
These facts suggest the investigation of a knot model in which the knot is
characterized solely by the algebra of the knot, $SU_q(2)$, independent
of its particular field realization.  If 
this model of the elementary particles has physical content,
one would expect the simplest particles to be the simplest knots (trefoils).
Consistent with this idea is the fact that there are four trefoils and
four classes of elementary fermions.  Moreover there is a unique
correspondence between the four trefoils and the four classes that is made
possible by the structure of the knot on the one hand and on the other hand
by the spectrum of charges $(0,-e,2/3e,-e/3)$ displayed by the four
classes of fermions.${}^3$  This correspondence is established
in Ref. 3 and stated in (2.2).

It is a simple formal extension of the conjecture that the trefoils are 
realized
as the lowest fermionic representations of the Lorentz group, to assume
that they are also realized as the lowest bosonic representations of the
Lorentz group.  Then there would be four Higgs-like scalars and suggests
that the search for scalars be widened.  

If these
particles are associated with minima of a Higgs-like potential, 
the potential
would have to be defined over the $SU_q(2)$ algebra.  One may define such
a potential, for example, by
\[
\langle 0|V(\bar\varphi\varphi)|0\rangle
\]
where $|0\rangle$ is the ground state of the $q$-algebra. 
This expectation value 
may be so defined that the minima lie at the trefoil
points of the algebra, i.e., at the four points represented by $\varphi =
D^{N/2}_{\frac{w}{2}\frac{r+1}{2}}$ where $(w,r)$ describes a trefoil.

The present model depends on two parameters, $q$ and $\beta$.  In comparing this
model with empirical data, such as the mass ratios of the vector bosons or
the mass ratios of the elementary fermions, the values of $q$ and $\beta$
have been found to
fluctuate.${}^2$  That is to be expected since the knot model 
as here presented is based on only the standard electroweak theory 
and ignores gluon and gravitational 
interactions.  The parameters, $q$ and $\beta$, may be regarded as
surrogates for the neglected interactions or as simply the parameters of an
effective Lagrangian.

Since the model is characterized by the realization of the trefoil as either
of the two lowest representations of the Lorentz group, namely the chiral
fermion and the scalar boson, a supersymmetric extension is naturally
suggested.

\vskip.5cm

\no {\bf References.}

\begin{enumerate}
\item R. J. Finkelstein, Int. J. Mod. Phys. A{\bf 20}, 487 (2005).
\item R. J. Finkelstein and A. C. Cadavid, Int. J. Mod. Phys.
A{\bf 21}, 4269 (2006).
\item R. J. Finkelstein, ``The Elementary Particles as Quantum Knots
in Electroweak Theory", hep-th/0705.3656.
\end{enumerate}

\end{document}